\documentclass[prb, reprint, superscriptaddress]{revtex4-1}

\usepackage{graphicx}
\usepackage{dcolumn}
\usepackage{bm}
\usepackage{here}

\begin{document}
\title{Structural and electrical characterization of the monolayer Kondo-lattice compound CePt$_6$/Pt(111)}

\author{Koichiro Ienaga}
\email{ienaga.k.aa@m.titech.ac.jp}
\affiliation{Institute for Solid State Physics, The University of Tokyo, Kashiwa, Chiba 277-8581, Japan}
\affiliation{Department of Physics, Tokyo Institute of Technology, Meguro-ku, Tokyo 152-8551, Japan}
\author{Sunghun Kim}
\affiliation{Institute for Solid State Physics, The University of Tokyo, Kashiwa, Chiba 277-8581, Japan}
\affiliation{Department of Physics, Korea Advanced Institute of Science and Technology, Daejeon 34141, Republic of Korea}
\author{Toshio Miyamachi}
\affiliation{Institute for Solid State Physics, The University of Tokyo, Kashiwa, Chiba 277-8581, Japan}
\affiliation{Institute of Materials and Systems for Sustainability, Nagoya University, Nagoya, Aichi 464-8601, Japan}
\affiliation{Graduate School of Engineering and School of Engineering, Nagoya University, Nagoya, Aichi 464-8603, Japan}
\author{Fumio Komori}
\affiliation{Institute for Solid State Physics, The University of Tokyo, Kashiwa, Chiba 277-8581, Japan}
\affiliation{Institute of Industrial Science, The University of Tokyo, Meguro-ku, Tokyo 153-8505, Japan}
\date{\today}
              
\begin{abstract}
We report the growth process, structure, and electronic states of 1 unit layer (u.l.) of a Ce-Pt intermetallic compound on Pt(111) using scanning tunneling microscopy/spectroscopy (STM/STS) and low-energy electron diffraction. An ordered $(2 \times 2)$ structure was observed in the form of films or nanoislands depending on Ce dose by annealing at around 700 K. A structural model constructed from atomically resolved STM images and quasiparticle interference (QPI) patterns indicates the formation of a new surface compound 1 u.l. CePt$_6$ on Pt(111) terminated by a Pt layer. A lateral lattice constant of the 1 u.l. CePt$_6$ on Pt(111) is expanded from the value of a bulk CePt$_5$ crystal to match the Pt(111) substrate. By measuring d$I$/d$V$ spectra and QPI, we observed an onset energy of the surface state found on Pt(111) above Fermi energy ($E_F$) shifts below $E_F$ on the Pt layer of the 1 u.l. CePt$_6$ due to charge transfer from the underneath CePt$_2$ layer. We discuss a possible two-dimensional coherent Kondo effect with the observed spectra on the 1 u.l. CePt$_6$.
\end{abstract}

\pacs{ }
\maketitle

\section{Introduction}
Studies of $f$-electron compounds have revealed many intriguing phenomena concerning strongly correlated electrons\cite{Hewson, Lohneysen2007, Yang2008, Gegenwart2008}. 
Hybridization of the localized $f$-electron states and itinerant electrons leads to the Kondo screening with a formation of the Kondo resonance state at low temperatures, which promotes an itinerant character of the $f$ electrons. With decreasing temperature further, an interplay between the $f$ electrons develops via hybridization, and causes the coherent Kondo state with a hybridization gap, an ordered state with the Ruderman-Kittel-Kasuya-Yosida (RKKY) coupling, or heavy-fermion superconductivity, depending on the strength of the hybridization. In the last decade, the surface-science techniques with high energy and spatial resolutions in real and reciprocal spaces have provided new insight into the novel electronic states\cite{Morr2016, Kirchner2020} using bulk single crystals\cite{Davis2010, Ernst2011, Yazdani2012, Park2012, Yazdani2013, Zhang2013, Yoshida2017, Miyamachi2017, Wahl2018, Giannakis2019, Shiga2021} and epitaxially grown films\cite{Haze2018}.

In the studies of bulk heavy-fermion compounds, the hybridization has been tuned by external pressure and element substitution. Recently, applications of nanoscale structural control to the strongly correlated systems have stimulated further development of the field\cite{Shishido2010, Coleman2010_news, KondoSuperlattice_review, Tsukahara2011, ienaga2012, Kawakami2013, Lohneysen2017, Raczkowski2019, Manoharan2019, Bachmann2019}. 
A pioneering work of the epitaxially grown (CeIn$_3$)$_m$/(LaIn$_3$)$_n$ artificial superlattice demonstrated a dimensional control of the heavy-fermion compound\cite{Shishido2010, Coleman2010_news}. With decreasing the layer number $m$, the antiferromagnetic (AFM) order observed in bulk CeIn$_3$ is gradually suppressed, the effective mass of electrons increases, and finally resistive behavior suggesting quantum criticality appears at $m$ = 2. 

Furthermore, a lateral-size control of a two-dimensional (2D) Kondo lattice is an exciting platform to study a spectral evolution from the single-impurity Kondo effect to the Kondo lattice behavior using scanning tunneling microscopy/spectroscopy (STM/STS)\cite{Tsukahara2011, Raczkowski2019, Manoharan2019}. So far, the platforms were provided by self-assembled clusters of Fe-phthalocyanine on Au(111)\cite{Tsukahara2011} and periodically arranged Co adatoms on Cu(111) with atomic manipulation\cite{Manoharan2019}. 
In the former system, a diplike Fano-Kondo resonance obviously splits at the center of the lattice, which was interpreted as AFM RKKY coupling\cite{Wahl2007} or the SU(4) Kondo effect\cite{Minamitani2012, Lobos2014} rather than a hybridization gap. 
In the latter, a diplike resonance is broadened at the center of the lattice, which was interpreted as an evolution into a hybridization gap\cite{Manoharan2019}. 
Therefore, the 2D coherent Kondo effect in the $f$-electron system, including comparisons with these, is even more important.  

An interesting example of the nanoscale structural control is CePt$_5$, which consists of alternating layers of a Pt$_3$ kagome lattice and a CePt$_2$ hexagonal lattice. The bulk CePt$_5$ has a resistivity minimum indicating the Kondo effect at about 10 K \cite{Sagmeister1997} and subsequently undergoes an AFM transition at 1.0 K \cite{Lohneysen1988}, exhibiting no signs of the coherent Kondo effect. In contrast, multilayered thin films of CePt$_5$ grown on Pt(111) exhibit features of the coherent Kondo effect, such as a tail of the Kondo resonance peak enhanced at a low temperature\cite{Garnier1997_PRBR, Garnier1998}, the $f$-electron spectral function with strong hybridization\cite{CePt5_ARPES_2}, and the formation of a hybridization gap\cite{CePt5_ARPES} in angle-resolved photoemission spectroscopy (ARPES) measurements. From these experiments, the Kondo temperature $T_{K} \sim$ 100 K and the coherence temperature $T^* \lesssim$ 20 K were derived in the multilayer CePt$_5$/Pt(111). The thickness-dependent electronic properties of CePt$_5$/Pt(111) were also reported with x-ray photoemission spectroscopy, x-ray absorption spectroscopy, and magnetic circular dichroism\cite{Tang1993, Fauth2015, Fauth2017}. The marked contrast between the bulk and the thin film could be explained by the reduced dimensionality\cite{Shishido2010}. 

Several studies reported compositions and thickness-dependent structures of Ce-Pt surface compounds on Pt(111) using STM and low-energy electron diffraction (LEED)\cite{Tang1993, Baddeley1997, Berner2002, Essen2009, Bode2014, Fauth2015, Fauth2015_2, CePt5_ARPES_2}. As summarized comprehensively in Refs. \onlinecite{Bode2014, Fauth2015}, these results contain discrepancies especially in the observed superstructures, but share some general trends as follows. 
First, CePt$_5$ is preferentially formed among possible compositions with the Pt-terminated surface. A recent theoretical calculation predicted that kagome holes of the surface Pt$_3$ layer are filled by additional Pt atoms, forming a chemically inert Pt$_4$ surface\cite{Tereshchuk2015}. This prediction was confirmed by LEED analysis at a 4 unit layer (u.l.) film\cite{Fauth2015_2}, and a similar structure was suggested for other thicknesses\cite{Fauth2015}. 
Second, the lateral lattice constant is a few percent smaller than that of the Pt(111) surface for $\geq$ 2 u.l. and approaches that of bulk CePt$_5$ with increasing thickness.

However, the growth of 1 u.l. films and nanoislands still remains highly controversial. For the 1 u.l. film, three structures have been reported with LEED measurements so far: (i) a weak (2 $\times$ 2) structure superimposed on a (5.6 $\times$ 5.6)$R30^{\circ}$ structure\cite{Baddeley1997, Berner2002}, (ii) an ideal (2 $\times$ 2) structure\cite{Essen2009, CePt5_ARPES_2}, and (iii) a faint (2 $\times$ 2) structure deriving from a long-ranged superstructure\cite{Bode2014, Fauth2015}. The structures (i) and (iii) were also confirmed by STM measurements, and an atomically resolved image was clearly obtained on (i). All the surface terminations were indicated to be a Pt layer by atomically resolved STM images and reactivity toward oxygen gases on (i)\cite{Baddeley1997}, CO absorption experiments with high-resolution electron energy loss spectroscopy on (ii)\cite{Essen2009}, and LEED $I$-$V$ analysis on (iii)\cite{Fauth2015, Fauth2015_2}. 
As for nanoislands, a STM study reported formation of CePt$_2$ islands with a (2 $\times$ 2) pattern\cite{Berner2002}. Another LEED study stated that annealing a sample at low Ce coverages restores a sharp (1 $\times$ 1) pattern of the substrate, indicating the absence of island formation of a Ce-Pt compound\cite{Essen2009}. These contradictory results need to be further investigated by an atomically resolved observation.
Furthermore, spectroscopic studies using STS or ARPES are still lacking for the 1 u.l. CePt$_5$/Pt(111) in contrast to the ARPES studies for the multilayer CePt$_5$/Pt(111). 
The 1 u.l. film has a purely 2D Ce lattice, and is suitable for the study of the 2D coherent Kondo effect.
Demonstrating the heavy-fermion behavior in such ultrathin films is necessary to further develop the fascinating fields that cross nanoscience and strongly correlated systems.
 
In the present paper, we study the growth of 1 u.l. films and islands of a Ce-Pt intermetallic compound on Pt(111). We observed the sample surfaces using STM and LEED in each growth step including a clean Pt(111) substrate, just after depositing Ce atoms, and after subsequent annealing. As-deposited dendritic morphology changes into a flat film with an ordered (2 $\times$ 2) structure by annealing around 700 K. 
An atomically resolved STM image of the (2 $\times$ 2) structure is well reproduced by a structural model of a 1 u.l. film of CePt$_6$ on Pt(111), not CePt$_5$, where the surface is terminated by a chemically inert Pt$_4$ layer consisting of a Pt$_3$ kagome lattice and additional Pt atoms onto the kagome holes, nominally forming a new surface compound. 
The lattice constant is expanded from the value of bulk CePt$_5$ to match the Pt(111) substrate. These results agree with the previous LEED images\cite{Essen2009, CePt5_ARPES_2}, the theoretical prediction\cite{Tereshchuk2015}, and the structure proposed based on the LEED $I$-$V$ analysis\cite{Fauth2015_2}.  
At low Ce coverages, we also found formation of nano-islands with the same structure as the 1 u.l. CePt$_6$ film. This finding paves the way for a lateral-size control of a 2D Kondo lattice\cite{Raczkowski2019}.
Furthermore, we measured d$I$/d$V$ spectra on the clean Pt(111) surface and the 1 u.l. CePt$_6$ film. On the Pt(111) surface, a well-known unoccupied surface state\cite{Pt(111)STS, Pt(111)STS_2} with an onset energy of 0.3 eV above Fermi energy ($E_F$) was confirmed. On the 1 u.l. CePt$_6$ film, a surface state similar to that on Pt(111) was observed, where the onset energy shifts down to $-$0.1 eV below $E_F$ due to charge transfer from Ce atoms. Finally, we discuss a possibility of the 2D coherent Kondo effect with spectra observed on the 1 u.l. CePt$_6$ film.

\section{Experimental Setup}\label{Sc:2}
A Ce-Pt compound was grown on Pt(111) in an ultrahigh-vacuum chamber, and then transferred {\it in situ} to a STM chamber. Base pressure of both chambers is below 1.0 $\times$ 10$^{-10}$ Torr. A clean and flat Pt(111) surface was prepared by several cycles of Ar$^{+}$ ion sputtering and annealing at 1200 K with the electron bombardment technique. Ce atoms were deposited onto the clean Pt(111) surface at room temperature (RT) using an alumina crucible. 
According to the previous reports\cite{Baddeley1997, Bode2014, Fauth2015}, an interatomic distance of deposited Ce atoms is 1.4 times larger than that of the Pt(111) substrate, leading to one Ce atom per two Pt(111) surface unit cells. In this paper, we define 1 ML as the number of atoms per area in the Pt(111) substrate and thus a monoatomic Ce film covering the entire substrate corresponds to 0.50 ML Ce. 
The sample was annealed around 700 K for 15 min after depositing Ce atoms to obtain an ordered Ce-Pt compound. Assuming the formation of CePt$_5$ as a possible composition, 0.25 ML Ce (apparent coverage 50 \%) on Pt(111) corresponds to a 1 u.l. film consisting of a Pt layer and a CePt$_2$ layer (see Fig. \ref{fig:4}). The optimum annealing temperature to obtain a well-ordered structure changes from 680 K for 0.10 ML Ce to 720 K for 0.25 ML Ce as mentioned in Sec. \ref{Sc:B} and Sec. \ref{Sc:D}.  

All the STM/STS measurements were performed at 5 K with a tungsten tip. To obtain topographic images, we used a constant tunneling-current ($I_t$) mode at a fixed sample bias voltage ($V_b$). The tunneling spectra d$I$/d$V$ were recorded using a standard lock-in technique with a modulation voltage of 719 Hz. The tip-sample distance was stabilized before sweeping the bias voltage using initial set-point values of bias voltage ($V_s$) and tunneling current ($I_s$). 

\begin{figure}[pbt]
	\centering
	\includegraphics*[width=8.6cm]{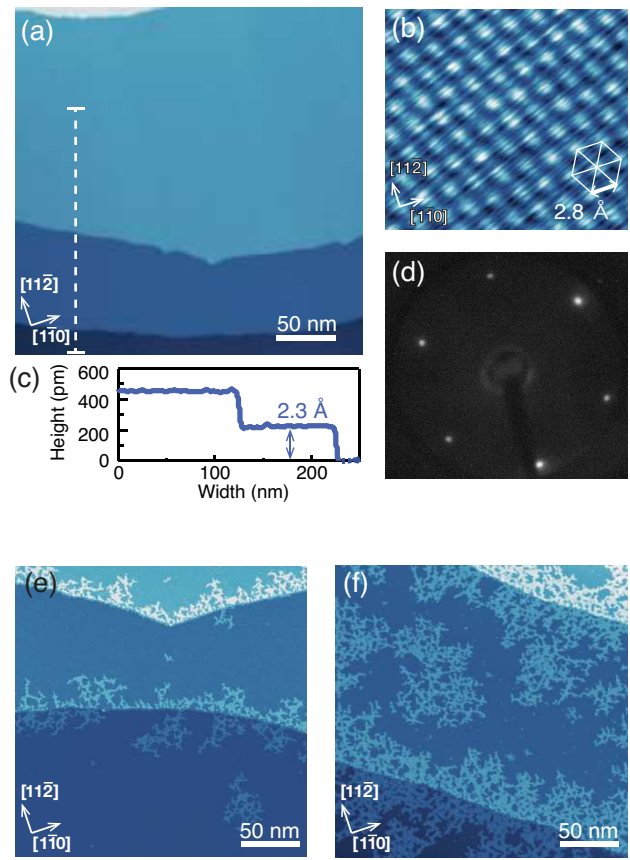} 
	\caption{\label{fig:1} (a) A topographic image of a clean Pt(111) surface at $V_b$ = 2.0 V and $I_t$ = 0.1 nA. (b) An atomic image of the Pt(111) surface at $V_b$ = 20 mV and $I_t$ = 50 nA. The white hexagon indicates an interatomic distance of 2.8 $\rm \AA$. (c) A height profile along the dashed white line in (a). (d) A LEED image of the Pt(111) substrate at 90 eV. (e),(f) Large-scale topographic images of as-deposited islands on Pt(111) surfaces exposed by (e) 0.05 ML and (f) 0.20 ML Ce atoms at $V_b$ = 1.0 V and $I_t$ = 0.1 nA. The islands exhibit a dendritic shape. All STM images (a), (b), (e), and (f) were obtained at 5 K.}
\end{figure}

 \begin{figure}[tbp]
	\centering
	\includegraphics*[width=8.6cm]{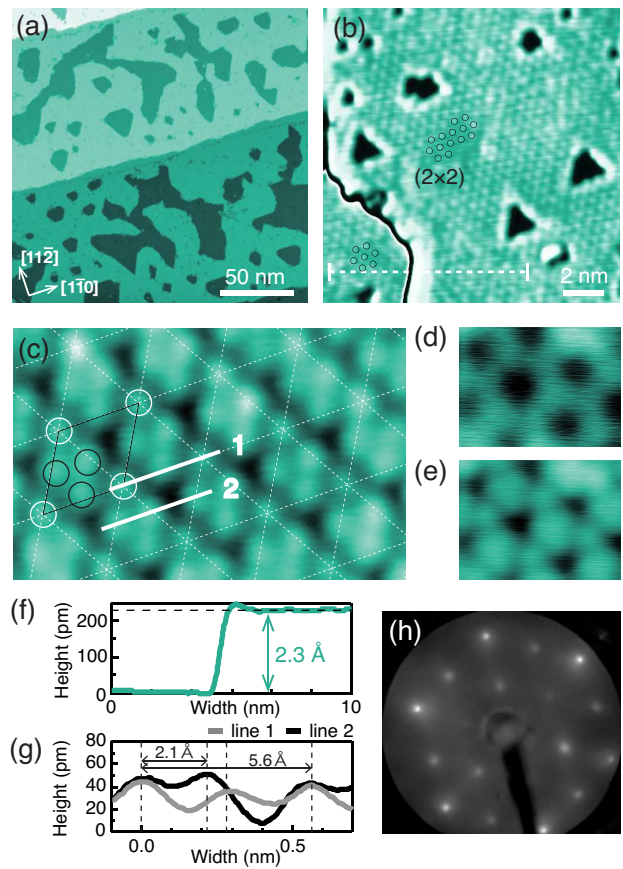}
	\caption{\label{fig:3} (a),(b) Large scale topographic images of a 1 u.l. Ce-Pt compound obtained by depositing 0.25 ML Ce on a Pt(111) substrate and annealing at 720 K for 15 min, which were scanned at $V_b$ = 1.0 V with (a) $I_t$ = 0.1 nA and (b) $I_t$ = 1.0 nA. Black circles in (b) indicate a $(2 \times 2)$ superstructure. The contrast of the image is artificially enhanced in (b) for clarifying the $(2 \times 2)$ protrusions. (c) An atomic image of the 1 u.l. Ce-Pt compound at $V_b$ = 30 mV, $I_t$ = 120 nA. The $(2 \times 2)$ superstructure is clearly identified with a unit cell indicated by a rhombus, which contains a hexagonal lattice and a trimer marked by white and black circles, respectively. (d),(e) Images of the $(2 \times 2)$ superstructure obtained at $V_b$ = 30 mV with (d) $I_t$ = 0.5 nA and (e) $I_t$ = 20 nA, respectively. All STM images (a)-(e) were obtained at 5 K. (f) A height profile along the dashed white line in (b). (g) Height profiles along solid white lines in (c). (h) A corresponding LEED image at 90 eV exhibiting $(2 \times 2)$ diffraction spots with a threefold symmetry.}
\end{figure} 

\begin{figure*}[tbp]
	\centering
	\includegraphics*[width=15cm]{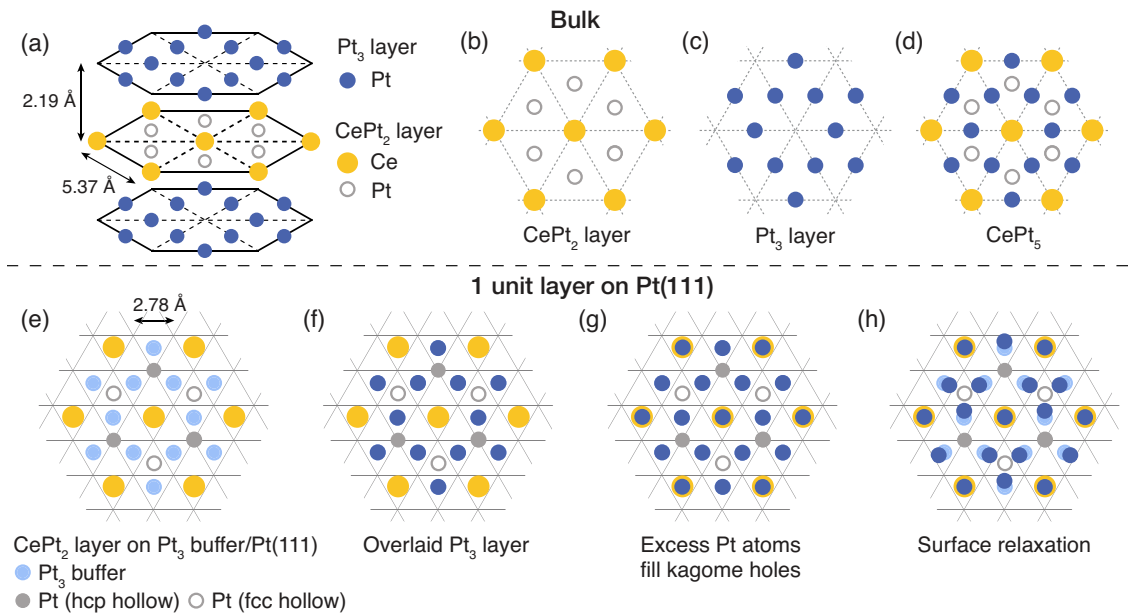}
	\caption{\label{fig:4} (a)-(d) Schematics of bulk CePt$_5$. The crystal structure is depicted in (a). Blue and yellow filled circles and gray open circles correspond to Pt atom in a Pt$_3$ kagome layer and Ce and Pt atoms in a CePt$_2$ layer, respectively. (b)-(d) correspond to the CePt$_2$ layer, the Pt$_3$ kagome layer, and the superposition of them, respectively. (e)-(h) Structural models of the 1 u.l. Ce-Pt compound on Pt(111). Solid black lines represent an underlying hexagonal lattice of Pt(111) surface. Inserting a Pt$_3$ buffer layer between the CePt$_2$ layer and the Pt(111) substrate is energetically favorable to stabilize the CePt$_2$ layer for Pt(111) as shown in (e), where Ce atoms with the largest ionic radius occupy the kagome holes of the buffer layer\cite{Tereshchuk2015}. Pt atoms in the CePt$_2$ layer occupy the fcc hollow sites and the hcp hollow sites of the buffer layer as depicted by gray open and filled circles, respectively. On the CePt$_2$ layer, the Pt$_3$ kagome layer is stacked in the same manner as the bulk crystal as in (f). In the Pt$_3$ kagome surface, the kagome holes atop Ce atoms are expected to be filled by additional Pt atoms as in (g)\cite{Tereshchuk2015}. The experimentally observed structure in Fig. \ref{fig:3}(c) is reproduced in (h) when trimers are formed in the Pt$_4$ surface to relax an energy difference between the Pt atoms in the CePt$_2$ layer occupying the fcc and hcp hollow sites of the buffer layer.}
\end{figure*}

\section{Results and Discussion}
\subsection{As-deposited structure of Ce/Pt(111)}

Figure \ref{fig:1}(a) displays a topographic image of flat and wide terraces of Pt(111) obtained by repeating the sputtering and annealing cycles. An atomically resolved image is shown in Fig. \ref{fig:1}(b). Surface contaminants such as carbon segregation from bulk, a major contaminant in Pt\cite{Musket1998_Book}, and adsorbates from vacuum were much suppressed. A step height of 2.3 $\rm \AA$ shown in Fig. \ref{fig:1}(c) and an interatomic distance of  2.8 $\rm \AA$ indicated by the white hexagon in Fig. \ref{fig:1}(b) agree with previous reports. A LEED image of the Pt(111) substrate in Fig. \ref{fig:1}(d) shows clear $(1 \times 1)$ diffraction spots with a threefold symmetry. 

Figures \ref{fig:1}(e) and \ref{fig:1}(f) show morphology of as-deposited islands on Pt(111) surfaces exposed by (e) 0.05 ML and (f) 0.20 ML Ce atoms. The islands exhibit a dendritic shape in a similar fashion to Au/Ru(0001)\cite{Au_dendritic}. As seen in Fig. \ref{fig:1}(e), the islands are preferentially formed near step edges, suggesting that Ce atoms diffuse on a clean surface before forming islands. As the density of the deposited Ce atoms increases, islands form in the middle of a terrace as in the case of 0.20 ML Ce in Fig. \ref{fig:1}(f). 

\subsection{1 u.l. Ce-Pt compound}\label{Sc:B}

Figures \ref{fig:3}(a) and \ref{fig:3}(b) show topographic images of a surface obtained by depositing 0.25 ML Ce at RT and annealing at 720 K for 15 min. As shown in a large-scale image of Fig. \ref{fig:3}(a), the terraces of the substrate are mostly covered by overlayers. We found formation of a superstructure over the whole surface, namely on the overlayers and the terraces, as indicated by black circles in Fig. \ref{fig:3}(b). 

Magnified images of the superstructure with different tip-sample distances are shown in Figs. \ref{fig:3}(c) - \ref{fig:3}(e). The atomically resolved image in Fig. \ref{fig:3}(c) clearly identifies a $(2 \times 2)$ structure. A unit cell of the $(2 \times 2)$ structure denoted by a rhombus contains a hexagonal lattice marked by white circles with an interatomic distance of 5.6 $\rm \AA$ [Fig. \ref{fig:3}(g)] and a trimer marked by black circles. Since area density of the protrusions in the $(2 \times 2)$ structure is the same as that of Pt(111), the formation of the superstructure is attributed to the trimer. Indeed, the distance between the two neighboring protrusions indicated by solid white line 2 in Fig. 2(c) is 2.1 $\rm \AA$ [Fig. \ref{fig:3}(g)], smaller than the interatomic distance of 2.8 $\rm \AA$ in Pt(111), exhibiting the trimer formation. Note that the atomic arrangement of the $(2 \times 2)$ structure has a threefold symmetry, which is consistent with a LEED image in Fig. \ref{fig:3}(h) exhibiting $(2 \times 2)$ diffraction spots with a threefold symmetry. No long-ranged superstructures were observed in the STM images. These results indicate that 1 u.l. of a Ce-Pt compound with the $(2 \times 2)$ structure is epitaxially grown on the Pt(111) substrate with inheriting the threefold symmetry of the substrate. 

\subsection{Structural model}\label{Sc:C}
To reproduce the observed $(2 \times 2)$ structure, we construct a structural model for the 1 u.l. Ce-Pt compound on Pt(111) by recalling a crystal structure of bulk CePt$_5$. As depicted in Fig. \ref{fig:4}(a), the bulk CePt$_5$ crystal consists of alternating layers of a CePt$_2$ layer and a Pt$_3$ kagome lattice with a lattice constant of 5.37 $\rm \AA$, which are shown in  Figs. \ref{fig:4}(b) and \ref{fig:4}(c), respectively. Blue and yellow filled circles and gray open circles correspond to Pt atoms in the Pt$_3$ layer and Ce and Pt atoms in the CePt$_2$ layer, respectively. Figure \ref{fig:4}(d) is a top view of the crystal structure.

\begin{figure}[tbp]
	\centering
	\includegraphics*[width=8.6cm]{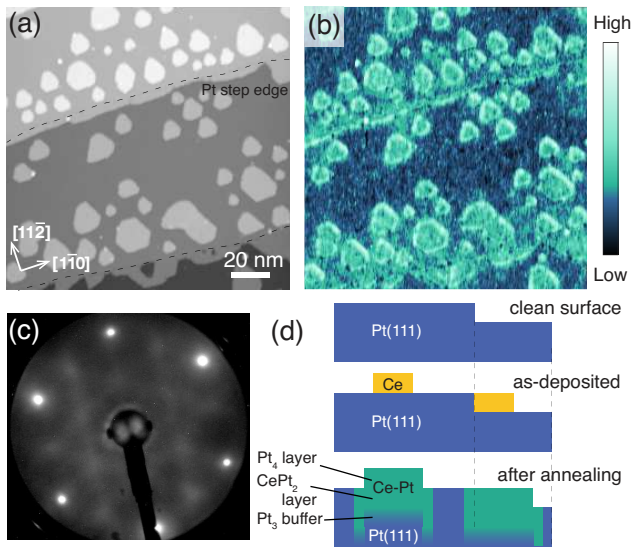}
	\caption{\label{fig:5} (a) A topographic image of a surface obtained by annealing at 680 K for 15 min after depositing 0.10 ML Ce, that was scanned with $V_b$ = 0.2 V and $I_t$ = 1.0 nA at 5 K. (b) A d$I$/d$V$ image obtained simultaneously with (a). (c) A corresponding LEED image at 90 eV. A weak $(2 \times 2)$ pattern with streaks connecting the spots was observed. (d) A growth model of the 1 u.l. CePt$_6$}
\end{figure} 

Next we consider a stacking arrangement of a 1 u.l. film of CePt$_5$ on the Pt(111) substrate. Solid black lines represent an underlying hexagonal lattice of the Pt(111) surface in Figs. \ref{fig:4}(e) - \ref{fig:4}(h). According to the theoretical calculation\cite{Tereshchuk2015}, inserting a Pt$_3$ buffer layer between the CePt$_2$ layer and the Pt(111) substrate is energetically favorable to stabilize the CePt$_2$ layer for Pt(111) as shown in Fig. \ref{fig:4}(e), where Ce atoms with the largest ionic radius occupy the kagome holes of the buffer layer. Pt atoms in the CePt$_2$ layer occupy the fcc hollow sites and the hcp hollow sites of the buffer layer as depicted by open and filled gray circles, respectively. In Fig. \ref{fig:4}(f), the surface Pt$_3$ kagome layer is stacked on the CePt$_2$ layer in the same manner as the bulk crystal for simplicity. In the Pt$_3$ surface, the kagome holes atop Ce atoms are expected to be filled by additional Pt atoms as in Fig. \ref{fig:4}(g). Here, the Pt$_4$ termination has been reported by theoretical calculation\cite{Tereshchuk2015} and LEED $I$-$V$ analysis\cite{Fauth2015, Fauth2015_2}. Furthermore, to relax an energy difference between the Pt atoms in the CePt$_2$ layer occupying the fcc and hcp hollow sites of the buffer layer, it is expected that trimers are formed around the Pt atoms on the fcc hollow sites as in Fig. \ref{fig:4}(h), where the observed $(2 \times 2)$ structure in Fig. \ref{fig:3}(c) is well reproduced. Thus, we conclude that the observed $(2 \times 2)$ structure originates from the 1 u.l. film of CePt$_5$ terminated by the `Pt$_4$' layer, where the nominal stoichiometry is CePt$_6$ due to excess Pt atoms filling the kagome hole. The topmost layer having the same number of atoms per area as Pt(111) probably results in the same step height of 2.3 $\rm \AA$ in Fig. \ref{fig:3}(f) as Pt(111). If the CePt$_2$ layer is directly overlaid on the Pt(111) substrate without the buffer layer, Pt atoms in the CePt$_2$ layer must occupy the hollow sites and the on-top sites of Pt(111). Even in this case, the protruding Pt atoms on the on-top sites will lead to the trimer formation in the Pt$_4$ surface.

The resultant structure with a clear threefold symmetry means that the 1 u.l. CePt$_6$ film is epitaxially grown on Pt(111) with a slightly larger lattice constant than that of bulk. This is consistent with a reported LEED pattern of the ideal (2 $\times$ 2)\cite{Essen2009, CePt5_ARPES_2} but inconsistent with the (2 $\times$ 2) + (5.6 $\times$ 5.6)$R30^{\circ}$\cite{Baddeley1997, Berner2002} and the faint (2 $\times$ 2) structure deriving from a long-ranged superstructure\cite{Bode2014, Fauth2015}. The latter two LEED patterns indicate that the 1 u.l. films are decoupled from Pt(111) as in the case of thicker CePt$_5$ films exhibiting a sixfold symmetry with long-ranged superstructure or rotated domains\cite{Baddeley1997, Essen2009, Bode2014, Fauth2015}. These controversial results indicate that there are several kinetic pathways to synthesize the metallic ultrathin layer, depending on growth parameters such as annealing temperature, deposition rate, amount of Ce, and surface cleanness. 

\subsection{Growth model}\label{Sc:D}

Figure \ref{fig:5}(a) displays a topographic image of a surface obtained by depositing 0.10 ML Ce at RT and subsequent annealing at 680 K for 15 min. There are hexagonal-shaped islands on the terraces. The surfaces of the islands show the same $(2 \times 2)$ structure as Fig. \ref{fig:3}(c), indicating the formation of the islands of 1 u.l. CePt$_6$ terminated by the Pt$_4$ layer. A corresponding LEED image in Fig. \ref{fig:5}(c) shows a weak $(2 \times 2)$ pattern. Weak streaks connecting the spots probably come from the island shape and defects inside the islands.  

Figure \ref{fig:5}(b) displays a d$I$/d$V$ image obtained simultaneously with Fig. \ref{fig:5}(a) at $V_b$ = 0.2 V, which is slightly smaller than an onset energy of a surface state of Pt(111)\cite{Pt(111)STS, Pt(111)STS_2}. As discussed in the next section, the color contrast in Fig. \ref{fig:5}(b) originates from a different onset energy of the Pt-terminated surface state due to carrier doping from Ce atoms. Thus, bright green areas in Fig. \ref{fig:5}(b) reflect the formation of the CePt$_6$. The bright green areas are confirmed not only at the islands on the terraces and at the ground layer surrounding the islands, but also along the step edges, indicating the formation of the extended islands along the step edges of the Pt(111) substrate denoted by dashed curves. Coverage of the bright area is approximately 40 \% of the whole image, which is consistent with the initial deposition of 0.10 ML Ce as explained in Sec. \ref{Sc:2}.

\begin{figure*}[!t]
	\centering
	\includegraphics*[width=17.2cm]{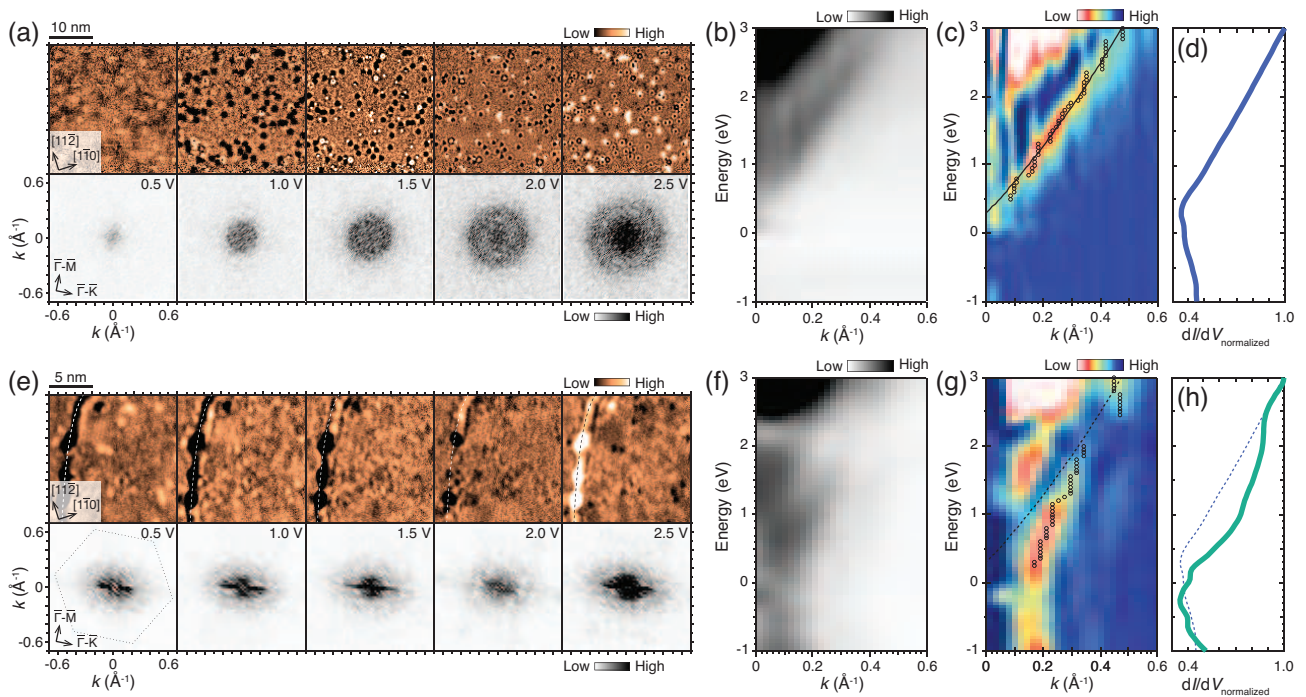}
	\caption{\label{fig:6} (a) Spatial mapping of d$I$/d$V$ spectra (top) and corresponding FT images (bottom) at selected values of the bias voltage obtained on bare Pt(111) at 5 K. The 128 $\times$ 128 spectra were recorded in a 30 $\times$ 30 nm$^2$ region. (Set points: $V_s$ = $-$1.0 V and $I_s$ = 0.5 nA; bias increment: 50 mV; modulation bias: 20 mV.) (b) Energy dependence of radially averaged profiles of FT images in (a). (c) Derivative of the profiles in (b) with respect to $k$. Open circles denote local minima reflecting surface-state dispersion of bare Pt(111) at each energy. A black curve fits the open circles as mentioned in the main text. (d) The spatial average of 128 $\times$ 128 d$I$/d$V$ spectra on Pt(111), which is normalized by a value at 3.0 V. 
(e) Spatial mapping of d$I$/d$V$ spectra (top) and corresponding FT images (bottom) at selected values of the bias voltage obtained on the 1 u.l. CePt$_6$ at 5 K. The 128 $\times$ 128 spectra were recorded in a 15 $\times$ 15 nm$^2$ region containing a step edge indicated by dashed white curves or a dashed black curve. A dotted hexagon in the bottom left panel shows the surface Brillouin zone calculated from the lattice constant, whose apexes agree with weak (2 $\times$ 2) spots observed in the FT images. (Set points: $V_s$ = 3.0 V and $I_s$ = 10 nA; bias increment: 50 mV; modulation bias: 20 mV.) (f) Energy-dependent profiles of FT images along the $\bar{\Gamma}$-$\bar{\rm K}$ direction perpendicular to the step edge. (g) Derivative of the profiles in (f) with respect to $k$. A dashed black curve in (g) is a copy of the black curve in (c). (h) The spatial average of 128 $\times$ 128 d$I$/d$V$ spectra on the 1 u.l. CePt$_6$. A dashed blue curve in (h) is a copy of the spectrum in (d).}
\end{figure*} 

From these results, we construct the growth model as in Fig. \ref{fig:5}(d). At RT, deposited Ce atoms diffuse on the clean Pt(111) terraces and preferentially settle near step edges forming the dendritic islands as shown in Figs. \ref{fig:1}(e) and \ref{fig:1}(f). By annealing, the 1 u.l. CePt$_6$ terminated by the $(2 \times 2)$ surface of Pt$_4$ is grown in the form of the hexagonal islands with the surrounding ground areas as in Fig. \ref{fig:5}(b). As the density of the deposited Ce atoms increases, the 1 u.l. CePt$_6$ area increases, and finally the whole surface is covered by the 1 u.l. CePt$_6$ as in Figs. \ref{fig:3}(a) and \ref{fig:3}(b).

\subsection{Electronic states: surface state}\label{Sc:E}

\begin{figure}[tbp]
	\centering
	\includegraphics*[width=8.6cm]{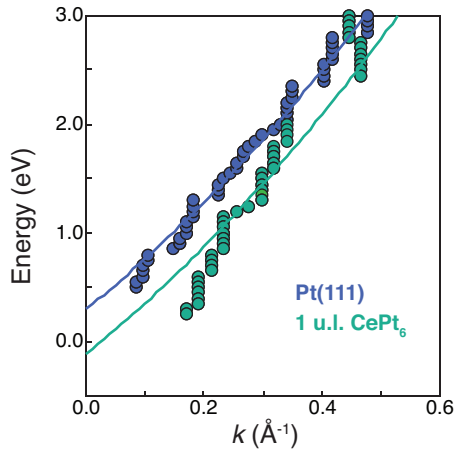}
	\caption{\label{fig:7} Extracted plots of the energy dispersion from Figs. \ref{fig:6}(c) and \ref{fig:6}(g). Blue and green circles are obtained on the Pt(111) and the 1 u.l. CePt$_6$, respectively. The blue circles are fitted by a blue curve using a quadratic polynomial function $E = E_0 + \alpha k + \beta k^2$ with $E_0$ = 0.3 eV. A green curve is drawn to fit green circles using the same fitting coefficients $\alpha$ and $\beta$ as the blue curve, resulting in $E_0 \sim$ $-$0.1 eV. }
\end{figure}

Next, we discuss the electronic state of the 1 u.l. CePt$_6$ by comparing with that of the Pt(111) surface. To investigate the band structures, the quasiparticle interference (QPI) was measured for the two samples. First we applied the QPI to Pt(111) to detect the well-known surface state as a reference. The upper panels of Fig. \ref{fig:6}(a) show real-space mapping of d$I$/d$V$ spectra measured on Pt(111). They are constructed by slicing d$I$/d$V$ spectra, which were recorded as a function of the sample bias voltage, at a selected bias voltage. The interference of standing waves induced by impurities or point defects causes spatial modulations of d$I$/d$V$ maps. The wave length $\lambda$ of the standing wave is governed by the scattering vector $q=2\pi/\lambda$ at a selected energy in the reciprocal space. With increasing the sample bias voltage, $\lambda$ decreases. This behavior is clearly seen in Fourier-transformation (FT) images in the lower panels of Fig. \ref{fig:6}(a). Here, we choose $k = \pi/\lambda (= q/2)$ [$\rm{\AA}^{-1}$] as the unit of the scattering vector in the FT images to compare with the previous work\cite{Pt(111)STS}. The radius of a ringlike pattern in FT images becomes larger with increasing sample bias voltage. Figure. \ref{fig:6}(b) exhibits the energy dependence of radially averaged profiles of the FT images, where the energy is converted from the bias voltage and the origin of the energy is $E_F$. Energy-dependent change of the dark area with high amplitude of the FT profile above 0.3 eV, namely energy dependence of the scattering vector, reflects the energy dispersion of the surface state of Pt(111) \cite{Pt(111)STS, Pt(111)STS_2}. A faint shadow at 0.1 eV near $k=0$ is identified as the contribution of the bulk $d$ band to the surface state\cite{Pt(111)STS}. To determine the energy dispersion of the scattering vector more clearly, the amplitude of the FT profile is differentiated with respect to $k$ as shown in Fig. \ref{fig:6}(c). 
The dispersion relation derived from a local minimum at each energy is denoted by open circles, which is quite consistent with the previous report\cite{Pt(111)STS}. A black curve is drawn by fitting the open circles as explained below. The spatially averaged d$I$/d$V$ spectrum on Pt(111) in Fig. \ref{fig:6}(d) also agrees with the previous work. A minimum at 0.3 eV and a slight hump at 0.1 eV correspond to an onset of the surface state and the contribution of the bulk $d$ band, respectively. 

The upper panels of Fig. \ref{fig:6}(e) show real-space mapping of d$I$/d$V$ spectra measured on the 1 u.l. CePt$_6$ containing a step edge indicated by dashed white or black curves, which is formed by annealing at 720 K for 15 min after depositing 0.25 ML Ce. They are depicted by slicing d$I$/d$V$ spectra as in Fig. \ref{fig:6}(a). Standing wave patterns generated by reflection at the step edge as well as the defects and impurities were observed. They result in elongated shadows perpendicular to the step edge direction in FT images in the lower panels. 
The surface Brillouin zone calculated from the (2 $\times$ 2) lattice constant is represented as a dotted hexagon in the bottom left panel, where the six apexes agree with weak (2 $\times$ 2) spots observed in all FT images. The elongated direction corresponds to the $\bar{\Gamma}$-$\bar{\rm{K}}$ direction because the step edge is formed along the nearest neighbor direction, leading to sensitive detection of the $\bar{\Gamma}$-$\bar{\rm{K}}$ dispersion. Figures. \ref{fig:6}(f) and \ref{fig:6}(g) show energy-dependent profiles of FT images along the $\bar{\Gamma}$-$\bar{\rm K}$ direction perpendicular to the step edge and their derivative with respect to $k$, respectively. An energy dispersion similar to that obtained on Pt(111) with a downward shift was confirmed as indicated by open circles. This dispersion is most likely attributed to the surface state of the topmost Pt$_4$ layer having the same number of atoms per area as Pt(111), which is shifted downward because of charge transfer from the underneath CePt$_2$ layer. The result gives further evidence that the observed (2 $\times$ 2) structure derives from the formation of 1 u.l. CePt$_6$.

We extract the energy dispersion of the scattering vector obtained on the Pt(111) surface and the 1 u.l. CePt$_6$ from Figs. \ref{fig:6}(c) and \ref{fig:6}(g) as blue and green circles in Fig. \ref{fig:7}, respectively. The obtained dispersion of Pt(111) is not reproduced by a parabolic function but a quadratic polynomial function $E = E_0 + \alpha k + \beta k^2$ as discussed in Ref. \onlinecite{Pt(111)STS}. A blue curve in Fig. \ref{fig:7} is drawn with $E_0$ = 0.3 eV, where $E_0$ is determined from the minimum in the averaged spectrum shown in Fig. \ref{fig:6}(d). This curve corresponds to the black curve in Fig. \ref{fig:6}(c) and dashed black curve in Fig. \ref{fig:6}(g). To evaluate the charge transfer effect more quantitatively, we fit the dispersion of the 1 u.l. CePt$_6$ by the quadratic polynomial function using the same $\alpha$ and $\beta$. As a result, we obtain $E_0 \sim$ $-$0.1 eV. Therefore, we conclude that the downward shift due to the charge transfer from Ce atoms is $\sim$ $-$0.4 eV. A slight deviation from the fitting curve in the low energy region will be discussed in the next section.

\begin{figure}[tpb]
	\centering
	\includegraphics*[width=8.6cm]{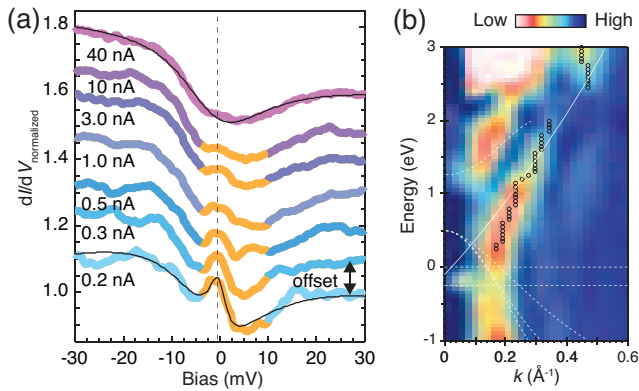}
	\caption{\label{fig:8} (a) High-resolution d$I$/d$V$ spectra near $E_F$ recorded at different set point currents $I_s$ and a fixed set point bias $V_s$ = 40 mV at 5 K. They are normalized by each value at 40 mV.  With increasing $I_s$, a double-dip structure around zero bias observed at $I_s$ = 0.2 nA gradually becomes broadened while keeping the peak position fixed as indicated by a dashed line, and finally changes into a single-dip structure at $I_s$ = 40 nA. The bias range where deviations from the single dip structure were observed is highlighted by orange. Black fitting curves at $I_s$ = 40 nA and 0.2 nA are drawn by the single and the two Fano curves as mentioned in the text, respectively. (Bias increment: 0.4 mV; modulation bias: 1 mV.) (b) Derivative of the FT profiles in Fig. \ref{fig:6}(f) with respect to $k$ [Fig. \ref{fig:6}(g)] is shown again, with the calculated spectral function based on the theoretical modeling of multilayer CePt$_5$ extracted partly from Ref. \onlinecite{CePt5_ARPES} as dashed white curves. Here, we adjusted the $\bar{\Gamma}$-$\bar{\rm{K}}$ distance of the calculation to the experimental value of the $k$-space distance between the origin and the (2 $\times$ 2) spots. A solid white curve is a fitting curve for an energy dispersion reflecting the surface state shown in Fig. \ref{fig:7}.}
\end{figure}

\subsection{Electronic states: possibility of the 2D coherent Kondo effect}\label{Sc:F}

Finally, we discuss the possible 2D coherent Kondo effect in the 1 u.l. CePt$_6$. 
High-resolution d$I$/d$V$ spectra near $E_F$ are shown in Fig. \ref{fig:8}(a), which were recorded at different set point currents $I_s$. Fano-like dip anomalies at zero bias were detected in all the spectra. With increasing $I_s$, a double-dip structure around zero bias observed at $I_s$ = 0.2 nA gradually becomes broadened while keeping the peak position fixed as indicated by a dashed line, and finally changes into a single-dip structure at $I_s$ = 40 nA. The bias range where deviations from the single dip structure were observed is highlighted by orange.

To discuss the possibility of the coherent Kondo state, we briefly review previous spectroscopic studies of the Kondo lattice systems.
In the Kondo lattice systems, the Kondo screening on the individual ions at $T \lesssim T_K$ leads to a formation of the Kondo resonance state at $E_F$. While ARPES directly detects the sharp state, STS and point contact spectroscopy (PCS) capture the resonance state as a Fano-Kondo resonance, where a quantum interference between a transfer into the Kondo resonance state and that into a conduction band leads to an asymmetric single dip or peak depending on the transfer ratio\cite{Schneider2008_review, Fano1961}. By further decreasing temperature, the hybridization gap is formed by band renormalization between the conduction band and the Kondo resonance state below $T^* (<< T_K)$\cite{Hewson, Lohneysen2007, Yang2008, Davis2010, Ernst2011, Yazdani2012, Park2012, Yazdani2013, Zhang2013, Yoshida2017, Miyamachi2017, Haze2018, Wahl2018, Giannakis2019, Shiga2021}. The detailed temperature dependence of the spectra throughout the overall process was reported using STS\cite{Ernst2011, Yazdani2012} and PCS\cite{Park2012, Zhang2013, Shiga2021}. In the case of the diplike Fano-Kondo resonance, which results from the tunneling process predominantly into the conduction band, the dip crosses over to a deep gap below $T^*$\cite{Yazdani2012, Zhang2013}.  In the case of the peaklike Fano-Kondo resonance due to strong coupling to the localized $f$ orbital, the peak splits by a gap growing below $T^*$\cite{Yazdani2012, Park2012, Shiga2021}. However, the hybridization gap is often broadened into a Fano-line shape showing an asymmetric peak\cite{Davis2010, Miyamachi2017, Giannakis2019} due to the lifetime effect of the heavy quasiparticles\cite{MDCmodel, MDCmodel2}. In some cases, the hybridization gap was incompletely observed as a peaklike anomaly inside the diplike Fano-Kondo resonance due to the possible momentum dependence of the hybridization\cite{Ernst2011}.

When we apply the interpretation of Ref. \onlinecite{Ernst2011} to the observed double-dip structures in Fig. \ref{fig:8}(a), they can be interpreted as an overlap of a wide Fano-like dip and a narrow peak corresponding to the Kondo resonance state and the incomplete hybridization gap. The latter can be also interpreted as the Fano-line shape owing to the smeared hybridization gap\cite{Davis2010, Miyamachi2017, Giannakis2019, MDCmodel, MDCmodel2}. The spectrum obtained at $I_s$ = 0.2 nA is well reproduced by a black curve using two Fano curves with a half width at half maximum (HWHM) of 11 meV for the wide dip and that of 2 meV for the narrow peak. Here, the single Fano curve is given by $f(\epsilon) = \frac{A}{1+q^2}\frac{(q+\epsilon)^2}{1+\epsilon^2}$, where $A$ is a fitting coefficient, $q$ a Fano factor, $\epsilon = \frac{eV-\epsilon_{0}}{\Gamma}$, $\epsilon_{0}$ the center of the Fano resonance, and $\Gamma$ a resonance width corresponding to the HWHM\cite{Schneider2008_review, Fano1961}. 
Our observation in the 1 u.l. CePt$_6$ is quite consistent with the previous ARPES study of multilayer CePt$_5$, where the Kondo resonance state with $T_K \sim$ 100 K and the hybridization gap of $\sim$ 2 meV were reported at 13 K\cite{CePt5_ARPES}. 
A similar $T_K$ was estimated from the temperature dependence of the Ce 4$f$ occupation number\cite{Fauth2015}. 
The spectrum obtained at $I_s$ = 40 nA in Fig. \ref{fig:8}(a) is fitted by the single Fano curve with a HWHM of 11 meV as a black curve. 
The smearing effect seems opposite to a simple expectation that a spectral feature becomes resolved well when $I_s$ increases, namely the tip approaches the surface.
The observed $I_s$ dependence of the spectra could be attributed to the lifetime effect of the heavy quasiparticles\cite{MDCmodel, MDCmodel2}.   
Although this effect has been considered to be due to impurity scattering, in the present ultrathin 1 u.l. CePt$_6$, a coupling between the tip and the sample possibly plays a more significant role than in bulk compounds\cite{MDCmodel2}.

The derivative of the energy-dependent profiles of the FT d$I$/d$V$ images obtained on the 1 u.l. CePt$_6$ [Fig. \ref{fig:6}(g)] is again shown in Fig. \ref{fig:8}(b) to discuss it from the viewpoint of the coherent Kondo effect. A flat dispersion just below $E_F$ around $k=0$ was obviously observed, while it is absent in Pt(111) as shown in Fig. \ref{fig:6}(c). This seems to connect with the dispersion curve below $E_F$ with a downward curvature not with that due to the surface state above $E_F$. 
The shape of the shoulder of the flat dispersion at $k \sim$ 0.2 $\rm{\AA^{-1}}$ is again consistent with ARPES data of the multilayer CePt$_5$, where the flat state corresponds to the Kondo resonance state and the hybridization gap of $\sim$ 2 meV was detected just above the shoulder\cite{CePt5_ARPES}. 

For further comparison, the calculated spectral function based on the theoretical modeling of multilayer CePt$_5$ is partly extracted from Ref. \onlinecite{CePt5_ARPES} onto Fig. \ref{fig:8}(b) as dashed white curves. The Kondo resonance state and its spin-orbit partner lie at $E_F$ and $-$0.25 eV, respectively, and the hybridization gap is formed at the intersections between these flat states and the conduction bands. Although our data are obtained on the 1 u.l. CePt$_6$, they resemble the calculation including an unoccupied band above 1 eV. At the present experimental resolution of 50 meV, the observed flat dispersion will include contributions from both the Kondo resonance state and the spin-orbit partner. 
A contribution from the low-$k$ part of the conduction bands may be hidden by the strong intensity of the dispersion due to the surface state, but its presence could cause the apparent downward deviation of the dispersion from the fitting curve (the solid white curve) mentioned in the previous section. A contribution from the flat band in the high-$k$ region is absent in our observation. This is probably because it mainly consists of the localized $f$-electron band\cite{Hewson, Yazdani2012}, which is basically difficult to access with STS, in contrast to the band near the hybridization gap containing conduction $spd$-electron components. Moreover, STS is sensitive to an electronic state near the $\bar{\Gamma}$ point. 
It is noted that we cannot rule out an interpretation that the bulk $d$ band near $E_F$ observed in Pt(111) shifts below $E_F$ in the 1 u.l. CePt$_6$. However, the spectral function for the multilayer CePt$_5$ did not report such $d$ band near $E_F$\cite{CePt5_ARPES}. 
We conclude that our results strongly indicate the 2D coherent Kondo effect in the ultrathin 1 u.l. CePt$_6$. For further discussion, theoretical calculations for the electronic states of the 1 u.l. CePt$_6$/Pt(111) are needed.

\section{Summary}
In summary, we studied the growth process, structure and electronic states of the 1 u.l. Ce-Pt intermetallic compound formed on Pt(111) using STM/STS and LEED. We observed that an as-deposited dendritic island changes into an ordered (2 $\times$ 2) structure by annealing around 700 K. The structural model constructed from the atomically resolved STM images of the (2 $\times$ 2) surface indicates the formation of 1 u.l. CePt$_6$ on Pt(111). The structural properties of the 1 u.l. CePt$_6$ such as the lattice constant, the surface termination, and the stacking relation were determined in an atomic scale, which are consistent with the reported LEED data \cite{Essen2009, CePt5_ARPES_2, Fauth2015_2}. At low Ce coverages, we also found the formation of nano-islands with the same structure as the 1 u.l. CePt$_6$ film. By measuring d$I$/d$V$ spectra and QPI, we found an energy dispersion due to the well-known unoccupied surface state on the clean Pt(111) surface with an onset energy of 0.3 eV\cite{Pt(111)STS, Pt(111)STS_2}. On the 1 u.l. CePt$_6$, the onset energy of the dispersion shifts down to $-$0.1 eV due to charge transfer from Ce atoms. This is consistent with the atomic structure model. 
Finally, we discuss the possible 2D coherent Kondo effect in the 1 u.l. CePt$_6$. We detected Fano-like anomalies around zero bias, which can be explained by the hybridization gap of $\sim 2$ meV and is consistent with the ARPES data of multilayer CePt$_5$/Pt(111)\cite{CePt5_ARPES}.
We also found a flat dispersion just below $E_F$ in QPI of the 1 u.l. CePt$_6$, which is absent in Pt(111). The observed shoulder shape of the flat dispersion again resembles the ARPES data of the multilayer CePt$_5$/Pt(111)\cite{CePt5_ARPES}, where the hybridization gap was observed above the shoulder. These results indicate the 2D coherent Kondo effect in the ultrathin 1 u.l. CePt$_6$.

\section*{ACKNOWLEDGEMENT}
We thank Masanobu Shiga, Tatsuya Kawae, Norikazu Kawamura, Takushi Iimori, Masamichi Yamada, and Yukio Takahashi for fruitful discussions. This work was partly supported by Grants-in-Aid for Scientific Research (KAKENHI) from the Japan Society for the Promotion of Science (Grant No. 26287061, 26790004, 18H01146, 19H02595, 20H05179, 20K14413, and 20K21119) and Basic Research Program through the National Research Foundation of Korea funded by the Ministry of Education (Grant No. 2019R1A6A1A10073887).


\end{document}